%% file: main.tex
\documentclass[sigconf]{acmart}
\setcopyright{rightsretained}
\copyrightyear{2023}
\acmYear{2023}
\acmDOI{XX.XXXXX/XXXXXXX.XXXXXXX}
\acmConference[Author pre-print]{Author pre-print, March 2023}{Chapel Hill, NC, USA}

\copyrightyear{2023} 
\acmYear{2023}

\usepackage{booktabs} 
\usepackage{subcaption}
\usepackage{mathtools}
\usepackage{multicol}
\usepackage{caption}
\usepackage{enumitem}
\usepackage{hhline}
\usepackage{url}
\usepackage{graphics}

\usepackage{graphicx}
\usepackage[font=scriptsize]{caption}
\usepackage{caption}

\usepackage{lipsum}

\begin{document}
\title[How does AI chat change search behaviors?]{How does AI chat change search behaviors?}

\settopmatter{printacmref=false}
\renewcommand\footnotetextcopyrightpermission[1]{} 
\fancyhead{}

\author{Rob Capra}
\affiliation{ 
    \institution{University of North Carolina at Chapel Hill}
    \city{Chapel Hill}
    \state{North Carolina}
    \country{USA}
}
\email{rcapra@unc.edu}

\author{Jaime Arguello}
\affiliation{ 
    \institution{University of North Carolina at Chapel Hill}
    \city{Chapel Hill}
    \state{North Carolina}
    \country{USA}
}
\email{jarguello@unc.edu}

\renewcommand{\shortauthors}{Capra}

\newcommand\todo[1]{\textcolor{red}{\textit{#1}}}
\newcommand\verify[1]{\textcolor{green}{\textit{#1}}}
\newcommand\cut[1]{\textcolor{blue}{\textit{#1}}}
\newcommand\propose[1]{\textcolor{magenta}{\textit{#1}}}

\begin{abstract}
Generative AI tools such as chatGPT are poised to change the way people engage with online information. Recently, Microsoft announced their "new Bing" search system which incorporates chat and generative AI technology from OpenAI. Google has announced plans to deploy search interfaces that incorporate similar types of technology. These new technologies will transform how people can search for information.
The research presented here is an early investigation into how people make use of a generative AI chat system (referred to simply as \emph{chat} from here on) as part of a search process, and how the incorporation of chat systems with existing search tools may effect users search behaviors and strategies.

We report on an exploratory user study with 10 participants who used a combined Chat+Search system that utilized the OpenAI GPT-3.5 API and the Bing Web Search v5 API.  Participants completed three search tasks.  In this pre-print paper\footnote{We are releasing this paper as an early pre-print version to disseminate results on this rapidly evolving topic and to be able to share prior to the ACM CHIIR 2023 conference.
\linebreak
\linebreak
Copyright 2023 Rob Capra and Jaime Arguello.}
of preliminary results, we report on ways that users integrated AI chat into their search process, things they liked and disliked about the chat system, their trust in the chat responses, and their mental models of how the chat system generated responses.

\vspace{-0.25cm}

\end{abstract}

\ccsdesc[500]{Information systems~Users and interactive retrieval}
\ccsdesc[300]{Human-centered computing~Human computer interaction (HCI)}

\keywords{Generative AI, chat systems, search interfaces, interactive information retrieval}

\maketitle

\input{paper}

\bibliographystyle{ACM-Reference-Format}


\end{document}

%% file: paper.tex
\section{Introduction}

Generative AI tools such as chatGPT are poised to change the way people engage with online information. Recently, Microsoft announced their "new Bing" search system which incorporates chat and generative AI technology from OpenAI. Google has announced plans to deploy search interfaces that incorporate similar types of technology. These new technologies will transform how people can search for information.

For years, the fields of information science, information retrieval, and human-computer interaction have studied how people make use of search systems and have investigated ways to improve search interfaces.  The integration of generative AI chat components is a major development that may profoundly change the way users interact with search systems.  Established knowledge about search behaviors and search interactions will need to be reconsidered and reevaluated in the context of these new tools.

The research presented here is an early investigation into how people make use of a generative AI chat system (refered to simply as \emph{chat} from here on) as part of a search process, and how the incorporation of chat systems with existing search tools may affect users' search behaviors and strategies.
To explore this, we conducted an exploratory user study with 10 participants. Our research questions were as follows:

\vspace{.1cm}
In the context of an information seeking task...
\begin{itemize}
    \item \textbf{RQ1.} How do users integrate chat into their search process?
    \item \textbf{RQ2.} Why do users engage with chat? 
    \item \textbf{RQ3.} What do users like about using chat?
    \item \textbf{RQ4.} What do users dislike about using chat?
    \item \textbf{RQ5.} Do user trust responses from chat?
    \item \textbf{RQ6.} What are users' mental models about how chat responses are generated and what they are based on?
\end{itemize}

Each participant in our study completed three search tasks in which they were asked to: 1) spend up to 20 minutes finding information using a Chat+Search system, 2) take notes they thought would be helpful, and 3) to record a short video explaining what they learned during their search.  The Chat+Search system combined a traditional search engine with an AI chat tool based on OpenAI's GPT-3.5 API. Participants were asked to think aloud during their search sessions.  We recorded the participants' screen and think-aloud comments during their search. After completing the tasks, we conducted semi-structured interviews with participants to learn more about their perceptions and use of the chat tool.

This study was conducted very quickly.  The study was approved by our IRB on February 15, 2023, data was collected from March 8 to March 15, and the data analysis presented in this paper was done from March 15 to March 18.  Our goal was to have a preprint paper ready to discuss with colleagues at the ACM CHIIR 2023 conference (March 19-23, 2023).  Given that the qualitative analysis was conducted by one author in a less than a week, the themes discussed in our results may not be comprehensive.  These are themes that we observed based on participants' actions, think-aloud comments, and interview responses.  Other themes may be present in the data but we may have missed them in this preliminary analysis.

At the time this study was conducted (March 2023), OpenAI's chatGPT had been publically available for a few months and had recently received significant attention in the press. AI chat components integrated with Bing\footnote{https://blogs.microsoft.com/blog/2023/02/07/reinventing-search-with-a-new-ai-powered-microsoft-bing-and-edge-your-copilot-for-the-web/} and Google's\footnote{https://blog.google/technology/ai/bard-google-ai-search-updates/} search products had been announced and released to limited sets of users, but had not been deployed to the general public.

We considered several options for how to run this study.  One was to simply ask participants to open a search engine in one window and chatGPT in another window and then give them information seeking tasks.  Another option was to attempt to get participants access to the ``new Bing'' interface that was in limited release at the time of the study.  We decided on a third route -- using the OpenAI GPT-3.5 API and the Bing Web Search v5 API to build our own combined "Chat+Search" system.  This option allowed us several advantages: 1) participants did not need to sign-up or have access to any limited release tools, 2) it allowed us to integrate the search and chat components, and 3) it allowed us to log user interactions (e.g., queries, clicks, mouse events, etc.) although analysis of that data is not presented in this paper.

\section{Related work}
In this early-release version of this paper, we do not discuss related work. However, there is extensive prior literature that has studied search behaviors, search interactions, search user interface design, search assistance tools, tools to support complex and exploratory searches, tools to support learning during search.  There is also extensive previous work on conversational AI and chatbots.  See papers in SIGIR, CHIIR, TOIS, CHI, JASIST, and IP\&M.

\section{Chat+Search System Design}
In this section, we describe the combined ``Chat+Search" system that we developed and used in this study. The system is shown in Figure~\ref{fig:chatsearch}.

\textbf{Task display.} The top-left of the screen displayed the scenario and task description for the current task.  Participants could refer to the description as needed while working on the task.

\textbf{WebSearch.}  The left side of the screen (see Figure~\ref{fig:chatsearch}) provided a ``traditional'' search engine interface (labelled ``WebSearch'') that included a textbox to enter a query and a display area to show a list of results.  Each result consisted of a title, URL, and query-biased text snippet.  Results were obtained using the Bing Web Search v5 API.  As in a traditional search engine, result links could be clicked to open the associated landing pages.

\textbf{ChatAI.} The right side of the screen (see Figure~\ref{fig:chatsearch}) provided a chat area labelled ``ChatAI'' that included a window that displayed the chat history and a textarea for the user to enter new questions/prompts to the chat (labelled ``Type anything here'').  When the user entered a prompt, it was appended to the conversation history and then the entire chat history was sent to the OpenAI completion API using the gpt-3.5 text-davinci-003 model with the following parameters:

\begin{verbatim}
    $open_ai->completion([
        'model' => 'text-davinci-003',
        'prompt' => $main_prompt,
        'temperature' => 0.9,
        'max_tokens' => 1000,
        'frequency_penalty' => 0,
        'presence_penalty' => 0.6,
    ]);
\end{verbatim}

While waiting for a response, a ``Processing...'' message was temporarily displayed at bottom of the chat.  When the response was returned, it was appended to the end of the chat.  We did not use OpenAI's streaming interface, so after submitting a prompt, users had to wait a few seconds for the chat response to be displayed.

The conversation history was the core of the prompt sent to the OpenAI API.  In order to create a friendly chatbox, we prepended the following instruction to the conversation history to form the complete prompt: ``Answer this as a chatbot that formats output in html.'' Typically, a response from the OpenAI API would return mainly text. The OpenAI API that we used did not provide source attribution links like the ``new Bing'' chat interface includes.

We wanted to create interactive connections between the chat and the WebSearch results. To do this, before displaying the chat response, we re-submitted the chat response to the OpenAI API with the following prompt prepended:

\begin{verbatim}
    List the most distinguishing noun phrases
    from the following text.
    Do not list more than 5 noun phrases.
    Do not list common words.
    Ouput the list in the form ["a", "b", "c", "d"].
\end{verbatim}

Then, in the final response shown to the user, we replaced each of the identified noun phrases with a clickable link that would issue that phrase as a new WebSearch.

\textbf{Queries were automatically copied to chat.} When a user submitted a query to the WebSearch interface, the system automatically copied the query and simultaneously submitted it as a prompt to the chat system (as if the user had entered it into the chat input area).  However, the reverse was not true.  When users entered prompts to the chat, the WebSearch results were not changed or updated in any way.

\begin{table*}[htp]
\fbox{
\includegraphics[width=16cm]{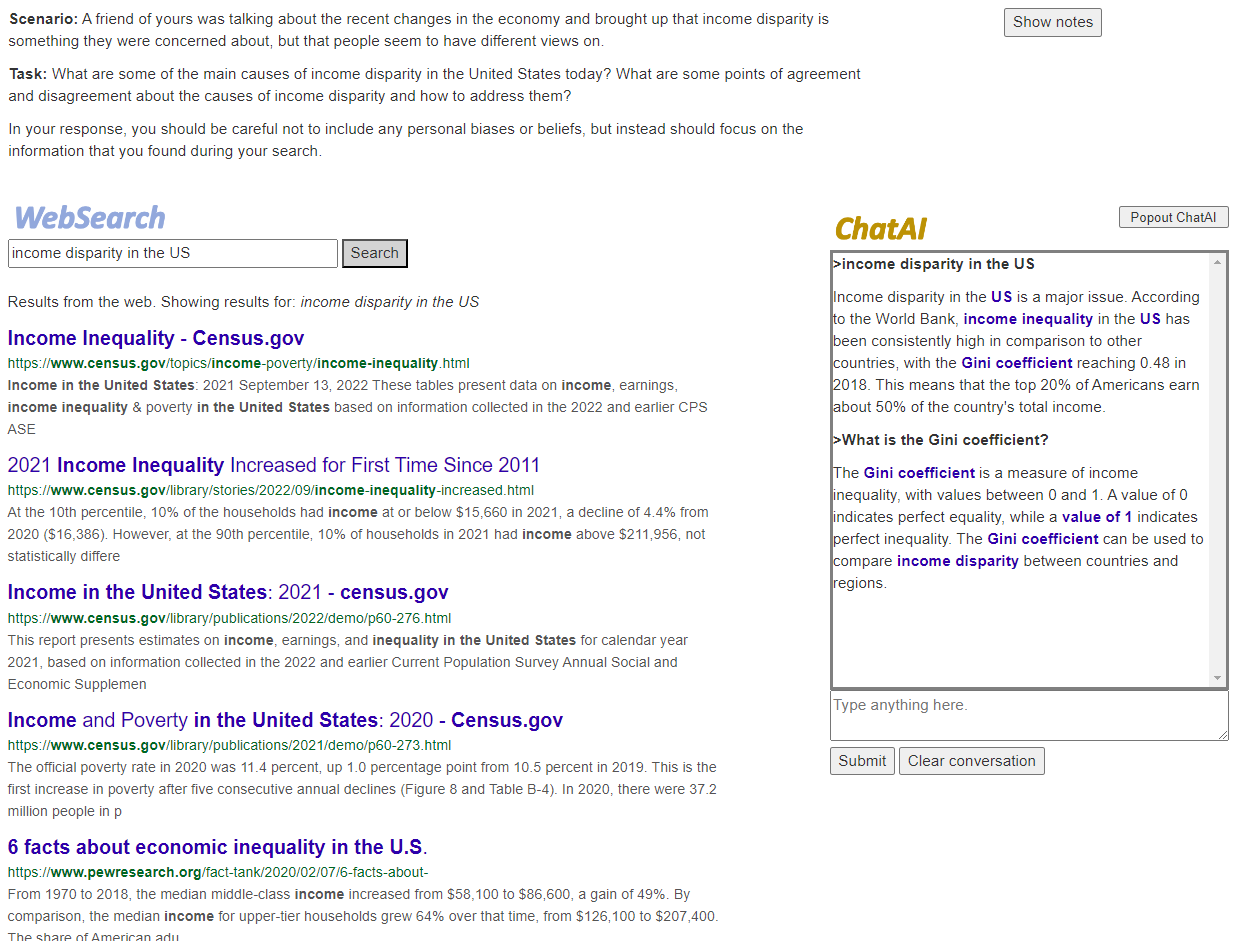}}
    \caption{Chat+Search interface.}
    \label{fig:chatsearch}
\end{table*}

\section{Method}
To investigate our research questions, we conducted a within-subjects user study with 10 participants. This study was reviewed by our university's Institutional Review Board and participants were shown information about what the study would involve before agreeing to participate.
\par

\subsection{Tasks}
\label{sec:tasks}
We asked participants to find information for three tasks: 1) to learn about osmosis and diffusion, 2) to determine the five best metrics to use to decide about investing in a stock, and 3) to learn about the causes of income disparity in the US and possible ways to address it.  We have used tasks 1 and 2 in prior studies of information seeking behaviors and learning.  Task 3 was designed to be a task on a topic of current debate.  The presentation order of the tasks was rotated across participants.\footnote{The three tasks result in 6 different orderings.  Since we had 10 participants, the rotations were not balanced.  For this exploratory study, we were not concerned about the slight imbalance in task presentation ordering.}

For all three tasks, participants were provided with a window to take notes in\footnote{The window used quill.js, a Javascript component for entering formatted text.} and were told that they would be asked to record a 2 minute video after their search to explain what they learned about the topic.  Prior studies have used this technique to successfully motivate participants to search for information on assigned tasks.

We situated each task in a fictional scenario (e.g. helping a family member).  This is a best-practice in interactive IR studies to help motivate participants and provide context for the information need. 

Each task also included the following instructions:
\begin{itemize}
    \item Your goal is to do a thorough search for information to address the task.
    \item You will have up to 20 minutes to search.
    \item As you search, you may take notes in the provided window.
    \item After you are done searching, you will be given 2 minutes to review your notes.
    \item Then, you will be asked to record a 2 minute video describing what you learned during the task.
\end{itemize}

We describe each of the three tasks below.

\subsubsection{Task 1: Learn about osmosis and diffusion}
For this task, participants were given the following scenario and task instructions:
\begin{quote}
    \textbf{Scenario}: One of your family members is a high school senior who is about to take an important biology exam.  Your family member has told you that she is struggling to understand the concepts of diffusion and osmosis and has asked for your help.

    \textbf{Task}: Your goal is to learn everything you can about the concepts of diffusion and osmosis.
\end{quote}

This task was designed to: 1) involve mainly objective information, 2) be open-ended and not well-defined in terms of the end goals, and 3) to require learning.

\subsubsection{Task 2: Metrics for stock investing}
For this task, participants were given the following scenario and task instructions:
\begin{quote}
    \textbf{Scenario}: You are thinking of investing in the stock market and want to compare different publicly traded companies based on different metrics. There are many metrics that can be used to evaluate whether a company is a good investment (i.e., is financially secure and well-positioned). 

    Metrics vary based on different criteria: (1) Is the metric easy to compute? (2) Is it easy to interpret? (3) Does it use information that is easily found on the web? and (4) Is it complementary (vs. redundant) when compared to the other chosen metrics?

    \textbf{Task}: In your opinion, what are the five best metrics to consider?  In this task, focus on metrics that are easy to compute using information freely available online.  For each metric explain: (1) how to compute it, (2) what it tells you, (3) what information you need to compute it, and (4) how the required information is easily found online.

\end{quote}

This task was designed to: 1) involve both objective and subjective information, 2) be well-defined in terms of the end goals, and 3) to require decision making (i.e., which are the \emph{best} metrics).

\subsubsection{Task 3: Income disparity in the US}
For this task, participants were given the following scenario and task instructions:
\begin{quote}
    \textbf{Scenario}: A friend of yours was talking about the recent changes in the economy and brought up that income disparity is something they were concerned about, but that people seem to have different views on.

    \textbf{Task}: What are some of the main causes of income disparity in the United States today?  What are some points of agreement and disagreement about the causes of income disparity and how to address them?

    In your response, you should be careful not to include any personal biases or beliefs, but instead should focus on the information that you found during your search.
\end{quote}

This task was designed to: 1) involve both objective and subjective information, 2) be defined, but not too specific in terms of the end goals, and 3) to require comparison and analysis.

\subsection{Participants}
Participants were recruited by sending a recruitment email to different academic departments around our university campus and asking them to distribute it to their students.  Using this method, we recruited 10 participants.  Participants ranged in age from 19 to 33 years  and self-reported their gender: female (9) and male (1). Participants were offered a USD \$50 Amazon gift card for participating.

\subsection{Protocol}
The study sessions were conducted remotely using the Zoom videoconferencing platform.  Each participant participated in an individual study session with the moderator.

Participants connected to the Zoom session and the moderator greeted them.  Next, the moderator asked if it was okay to record the session and sent the participant a link (via Zoom chat) to a ``main'' webpage used to guide the participant through the different parts of the study.  The participant then shared their screen with the moderator and opened a web browser to the main page.  Next, participants read an information sheet describing what would take place in the study and how their data would be protected.  If they agreed to proceed, they then filled out a short demographic questionnaire.  Then the moderator shared their screen and played a short ($\sim$1 minute) video introducing the combined Search+ChatAI system and demonstrating its main features.  After this, the moderator instructed the participant about the think-aloud protocol.

The participant then did the following steps for each of the three tasks.  First, they read the task scenario and description.  Next, they completed a pre-task questionnaire (not analyzed here).  Then, they opened the combined Search+ChatAI interface and worked on the task.  Participants were given up to 20 minutes to work on the task, but could stop sooner if they decided they had found enough information. The interface also included a pop-out window that could be used to take notes.  Participants were informed that when recording the video at the end of the task, they would have access to their notes, but not to any other windows. During the time the participant was searching, the moderator muted their sound and turned off their camera to avoid distracting the participant.  Also during the search sessions, the moderator took detailed notes about the participant's interactions with both the WebSearch and ChatAI components and the participant's search behaviors and strategies.

After finishing their searching, participants were given a few minutes to review their notes.  Then they were asked to close all other windows except the notes window and to record a brief ($\sim$2 to 3 minute) video to describe what they had learned during their search.  Finally, participants were asked to complete a post-task questionnaire (not analyzed here).

After completing all three tasks, the moderator conducted a semi-structured interview with the participant about their impressions and experiences doing the tasks.  During the interview, we asked the following questions:

\begin{itemize}
    \item How did the chat differ from using a traditional search engine?
    \item Did using the chat change anything about your search strategy?
    \item What did you like and dislike about the chat?
    \item What types of information did you use the chat to find and what types did you use the search to find?
    \item Did you have an idea about where the information from the chat was coming from or how it was generated?
    \item Did you trust the information from the chat?
\end{itemize}

At the end of the session, participants were emailed their Amazon gift card, asked to fill out a payment acknowledgment form, and thanked for their participation.

\section{Results}
In this section, we present a qualitative analysis of our observations of participants' search behaviors during the search sessions and of the participants' responses during the semi-structured interview.  For this early-release version of this paper, these qualitative analyses were performed only by the lead author.  Future work should confirm these results with additional qualitative coders and additional data. In the subsections below, we present our results organized by research question.
\par

\subsection{How do users integrate chat into their overall search process? (RQ1)}
One of the main questions we wanted to explore was how participants would incorporate chat into their overall search process.  Our participants were familiar with web search, but incorporating chat into this process was new to most of them (although a few had used chatGPT on a limited basis).

Based on our observations of the search sessions, we identified three main ways that participants interacted with chat as part of their overall search processes:  1) didn't use the chat tool, 2) used chat as a \emph{question-answering tool} to find answers to specific questions that arose during their search, and 3) used a \emph{chat-first approach} to get background information and lists of sub-topics to explore.

\textbf{Didn't engage with chat.}
Several of our participants either did not enter any questions to the chat, or only briefly looked at the chat responses that were generated automatically from their search queries.  For example, P9 was very proficient at web search -- quickly reviewing results, opening new tabs, and extracting information from web pages. P9 did not use the chat for any of the three tasks, and commented: ``I typically have an idea or a question in mind. And so I'm able to input that question into the search queries.''  Some participants also commented that they were used to web search and felt more comfortable with it. For example P5 described: ``It just feels more comfortable to use the system that is familiar [search].''

\textbf{Chat as a question answering tool.}
Chat was also incorporated into the search process as a question-answering tool.  Participants would switch to chat to get quick answers to specific questions that arose during their search.  For example, in the stock metrics task, P3 had done a web search, clicked a result, and was reading a landing page about a metric called ``total returns to shareholders''.  At first, this metric looked promising, but then P3 had doubts, commenting: ``sounds like a good thing... [after reading more]... maybe not".  At this point, P3 switched to chat and asked it: ``limitations of total return to shareholders in stock market.''

\textbf{Chat-first.}
Another common approach was for participants to start their search with chat in order to get background information, lists of important concepts/terms related to the topic, and definitions. For example, P7 described using a chat-first approach:  ``I think that the chat was really effective at summarizing information and giving me good places to, or almost like keywords to start looking for information.  It definitely gave me a platform that I would [be] able to go and do my other additional research with.''
P6 described using a similar approach for the stock metrics task: ``I would look up the best stock market metrics on the Chat AI, and it would give me a list. But then I would then individually look up those metrics... to go more in depth about what I was finding.''
And P1 commented: ``So I think it's [chat] really good for finding specific information quickly. But if you have to do through research, or you want to find different information. I would do the search after trying the chat function.''

\subsection{Why do users engage with chat? (RQ2)}
Participants engaged with the chat feature for a variety of different reasons including: as a starting point, for unfamiliar topics, to easily extract information, and because of time pressure.

\textbf{Chat as a starting point.}
As noted in the results for RQ1, many participants described how they used the chat as a starting point, even for topics they were familiar with.  For example, P1 noted: ``I think I used the chat to find simple definitions and things like that, or as a place to start.''
And P5 commented: ``I have a biology background, so that stuff that I already kind of knew. I just didn't really remember exactly, but it was like [I needed a] refresher... If the chat gives the right answers, I could... recognize whether or not it was telling... the wrong thing.''

\textbf{Chat was helpful for unfamiliar topics.}
Several participants described that chat was helpful for collecting and synthesizing information about unfamiliar topics.  For example, P3 described: ``I think it was most helpful for the first task, because that was the task that I knew the least about, and so, having it explained to me in layman's terms was definitely very helpful.'' And P1 described how chat saved effort for unfamiliar topics: ``I think it was better for things that I knew nothing about. It's a lot easier to start at the chat function to not have to look through a bunch of information that I know nothing about.''

\textbf{Chat was easier to extract information.}
Some participants noted how using the chat feature made it easier to extract information compared to doing so from a list of links returned by web search.  For example, P2 described: ``If you look up research articles [using search], you're basically just gonna get a bunch of different people's like, I studied this aspect of income disparity. And you're going to have to synthesize that.''

\textbf{Time pressure.}
Two participants described how the 20 minute time limit made them rely on the chat more than they normally would.  For example, toward the end of the stock metrics task, P5 was running low on time and noticeably shifted into a mode of relying more directly on the chat, in several cases copy/pasting the chat responses into their notes.

\subsection{What did participants like about chat? (RQ3)}
In the post-session interview, we asked specifically about what things participants liked about the chat feature.

\textbf{Concise, easy-to-understand answers.} Participants described liking that the chat provided simple, concise answers.  For example, P1 commented: ``It [chat] was a little more helpful when I was looking for quick information, because I could ask a question, and it would pull up one concise answer. As opposed to search where you have to filter through the answers, and maybe open an article and find the answer.'' Similarly, P5 described: ``It gave really quick summaries, which is always nice when you're asking questions.'' Finally, P2 noted: ``[For] the income disparity question basically I just asked it the questions that were asked to me, and it pretty much gave me the answers that I was expecting, because I was kind of familiar with the topics. I was like, oh, wow, okay. It really did answer the question almost in its entirety.''

\textbf{Synthesis and summary of information.} Participants commented that the chat feature provided synthesis and summaries of information, making it easier to digest.  P2 noted: ``I relied on it more and more because it was synthesizing the information for me.'' P6 specifically described how the chat synthesized information into short paragraphs: ``I did like how it synthesized a lot of information... I didn't have to go to different [websites]. Chat AI kind of synthesize[s] all of that into one small paragraph which I thought was pretty helpful.''  And P7 described how the synthesis provided by chat made their search process more efficient: ``I particularly like the way that I was able to.. type in a concept and get a quick response with things to look for.  I think it really sped up [the process]. Instead of having to look through a bunch of different sources for what I'm looking for, I was able to say one thing and refine it if I needed to.''

\textbf{Chat encouraged natural language questions.}
Some participants commented that it was easier to write natural language queries to the chat.
P1 noted: ``I think I formed more specific questions [with chat] because if I'm searching for something more broadly I'll just type in like one or two words, whereas it was easier to type in a specific question for the chat function.  I think it would maybe help with researching just to form questions to answer in terms of organizing.''

\subsection{What did participants dislike about chat? (RQ4)}
We asked participants about what things they disliked about the chat feature.

\textbf{Answers were too general.}
Several participants commented about how the chat tended to return answers that were too general in nature.  For example, P4 described: ``I didn't really like it just because it did give the generalized answer which I'm not a big personal fan of.'' Similarly, P6 noted: ``I like how that it synthesized it, but I think that's also kind of a con as well, because it's not as detailed, whereas like a website... you're more likely to get all of the details that you're trying to see.''

\textbf{Repetition in answers.} One participant described how the chat seemed to give similar answers, even when they reworded their question.  For example, P1 noted: ``[in search] rewording a questions is a lot easier to get a ton of different answers.  But then, with the chat, I got the around the same answer even if I change the wording of what I was asking."

\textbf{Contradictory answers.} One participant noted a confusing and seemingly contradictory answer provided by the chat. P7 recalled: ``It seemed there were a few times when it was sort of unclear. On the last example I asked it a question and it said no, but it sort of seemed to agree with me.''

\textbf{No videos, pictures, or other verticals}.  Participants also commented that the chat did not return videos, pictures, or other vertical search categories (e.g. shopping, news).  For example, P2 described: ``media like videos didn't come up as much as when you do a standard Google search.''  Our Chat+Search implementation relied on the Bing Web Search API and the OpenAI completion endpoint, but did not incorporate searches to other verticals.

\textbf{Search queries also sent to chat.}
For any query typed into the WebSearch query box, our Chat+Search system automatically sent a copy of the query as a prompt to the chat window.  Some participants explicitly made use of this feature to multi-task.
However, one participant felt that this automation created pressure to use the chat. P2 commented: ``When I typed things into the WebSearch, it also appeared in the [chat] AI, but it kind of felt like it was forcing me to use it.''

\textbf{Did not include links to sources.}
Many participant complained that the chat did not include links to sources to support its responses.
P2 described: ``That was where my skepticism was coming from because I wasn't I wasn't sure [of the sources]. You just can't know exactly what the sources are and what maybe you're missing out on. So I would use [chat] probably as a starting place, and then go way further into it.'' P4 noted: ``How do I know that that answer is right? I'm not seeing anything else to support it. So that would literally just be me taking this engine's word at face value rather than being able to check it out for myself.'' Finally, P5 described: ``Especially when I was trying to find out very factual information, it didn't let me see where those answers were coming from.''

We note that our chat implementation did not include specific links or references to sources. Links to sources were not returned from the OpenAI API that we used (although the completion endpoint occasionally included links as part of the main text of the response).  However, the chat incorporated into the ``new Bing'' interface available to a limited set of users in March 2023 does include links to sources/references for the chat responses.  Having an understanding of where information is coming from and links to go explore it on their own was clearly important to our participants.

\textbf{Links we provided in chat were not always helpful.}
Several participants noted that the links that we added to chat were not always helpful.  Recall that we extracted important noun phrases out of the chat response and created links to launch web searches for these phrases.  This approach had mixed results.  As P2 noted: ``Some of the links that are provided [by the chat] were good, but some of them were like... it underlined poverty... So I think maybe some of those it could improve on. Like maybe you want to read more about this, and then the link could be like studies have shown, and then it would send you to [those studies].'' And P4 said: ``The [links in the chat] don't really give like that much information. I would just take  specific words that I thought were important from the blurb and put them in the other search engine.''

\subsection{Do users trust chat? (RQ5)}
Participants gave a variety of responses to the question of whether they trusted the answers from the chat.

\textbf{I don't trust it.}
Some participants were very skeptical about the chat system.  For example, P5 did not trust the chat, saying: ``No because, I'm already a little skeptical about all those sources online... I have [also] had news stories pop up about how you shouldn't trust chat AI.''

\textbf{It sounded reasonable.}  Several participants described that the responses provided by the AI chat sounded correct, but that they were not completely sure if they were.  For example, P1 noted: ``For some of the answers I could discern and be like, oh, this like makes sense.  It seemed like everything was true information.''  And P2 described: ``Nothing sounded absolutely absurd or crazy or anything like that.''

\textbf{Don't trust, verify.} Some participants expressed a healthy level of skepticism about trusting the answers provided by the chat.  Several described that they would want to verify information coming from the chat.  For example, P2 commented: ``I would have definitely like to verify... especially when it said like studies have shown... I would want to see what those studies were and actually read them.''

\textbf{Trust depended on familiarity with the subject.}
For some participants, their level of trust in the chat responses depended on their prior level of knowledge about the topic.  For example, P3 said: ``[for the stock metrics task]... I was like I should probably check what the AI is telling me because I don't know anything about that. But in the income disparity, and [the] osmosis [task] it seems to agree with what I already knew.'' And P7 described: ``I think for the most part I did [trust the chat]. But there's almost something at the back of your  head that is sort of unsure.  So I think there were instances where I wasn't really too sure whether or not to trust it entirely. I think I had an easier time trusting it when it was affirming what I had already figured [out]. But if it was something I didn't know anything about... I wanted to confirm it with other sources.''

\textbf{Influence of interactions with other AI/chat systems.} 
One participant commented on how their prior use of other AI/chat systems influenced their view of the chat used in this study.  For example, P4 explained: ``I definitely didn't trust [chat] as much as the [search]. I know that, like Siri and Alexa and all of those... they mess up. You can ask them something, and they will tell you something completely unrelated. So knowing that... made me more inclined to trust my own judgment.''

\subsection{Mental models of chat (RQ6)}
We asked participants if they had an understanding of where the chat results were coming from or how they were generated.  Participants expressed varying levels of uncertainty about the origins of the chat responses.

\textbf{Unclear mental model.}
Some participants reported that they did not have a clear mental model of how the AI chat was generating its responses or what they were based on.  For example, when asked about where and how the chat answers were generated, P1 said: ``Not really.  I actually don't really know at all.'' Similarly, P6 said: ``I genuinely don't understand, like... I have no idea.''

\textbf{Responses based on previous search results.}
Several participants hypothesized that chat response were connected in some way to the WebSearch results, but some were unsure about the extent of the connection. In most of these cases, participants' beliefs that the chat response was connected to the WebSearch results were not correct or complete. For example, when asked about what the chat responses were based on, P2 hypothesized: ``Sometimes it was kind of clear, because, when I asked like investing strategies you could see that clearly. Some of those literally came from listicles\footnote{Listicles are articles written based on lists (e.g., an article describing the ``5 best stock metrics for beginning investors'').} that were [returned in the search results].'' P3 described: ``I assumed it was coming from the the same sort of resources that were popping up on the regular search engine, just sort of summarized.'' P4 speculated: ``It definitely seems like it kind of pulled some of the main ideas from the different websites that had popped up.'' And P5 wondered if it was similar to question answering technology they had seen before, describing: ``I assume [it is like] the thing Google has... if you ask a question, it will have a blurb answer. I think they usually lift that from some of the first sources [web results].''  In all these cases, participants seemed to express a belief that there was an explicit connection between the WebSearch results and the chat responses (i.e., the chat was summarizing the top WebSearch results).  While LLMs may have been trained on datasets that included the web pages, these speculations illustrate fundamental misunderstandings about how the LLM technology behind the AI chat works.

\textbf{Unsure what sources it has access to.}
Participants also described that they were not sure what sources the chat system had access to.  For example, P2 explained: ``I don't know if it has access to library articles that we would have. Is it just synthesizing information from open source material?'' And P7 noted: ``I know a little bit about chat. You know it draws from a huge... it's making patterns from a database of information, but it doesn't say here are the sources that I had to back this up.''

\section{Discussion}
Based on a preliminary analysis from this small-scale user study, we see: 1) promising potentials for generative AI to aid searchers, 2) reasons for concern that searchers will rely on biased or incorrect information, 3) opportunities for the interactive IR community to help design ways to more effectively incorporate generative AI into search systems, and 4) ideas to help educate searchers about how to (and how not to) use generative AI to search for information.  We elaborate on each of these below.

\textbf{Promising potentials.}
First, we observed many positive uses of AI chat to aid search processes.  Participants used chat to gain overviews of unfamiliar domains, to get quick answers to questions that arose during their search (thus avoiding complex subsearches), and using chat to help generate lists of starting points for their own web searches.  In addition, a majority of our participants expressed that they did not trust the output from the AI chat.  We observed many instances where participants used web search to do their own searches to confirm or learn more about topics returned in the chat responses. We view these as positive signs.  Largely, our participants used the AI chat as a helpful tool, but did not completely trust it.

We also note that in our research careers, we have studied many web search assistance tools. Anecdotally, we felt that the level of interaction with the AI chat was among the highest of any assistance tool we have studied.

However, a large caveat on these results is that our participants were all students from a large US university and clearly had high-levels of information literacy skills.  Even in their web searches, a majority of our participants considered, evaluated, and commented on the sources of information returned by the search engine, and many employed mature search strategies such as query revision, looking at multiple sources, and triangulation of results.  These critical information literacy skills appear to at least have partially transferred over into our participants' use of the chat AI for information seeking.  However, it is unclear how a broader population of searchers might or might not employ similar critical strategies.

\textbf{Reasons for concern.}
Despite our observations that participants in general did not trust the responses from the AI chat, there were many times that participants acted on the results from the chat in some way (e.g., by using it as part of web search terms, as a point of comparison against information found in web results, or directly copy/pasting the chat information into their notes).  This illustrates that the output of the AI chat had impacts on participants' search strategies and information-seeking approaches.  In situations where generative AI hallucinates or produces incorrect output, this could mislead or confuse searchers, or could require extra effort on the part of the searcher to check or verify the information. 

Another point of potential concern was the potential for \emph{trust transference}. Many participant described that their trust in the chat results depended on their prior familiarity and knowledge of the topic.  This has the potential to lead participants into a false sense of trust.  For example, if the AI chat returned three reasons for income disparity and the participant recognized two as being consistent with their prior knowledge, they might be inclined to believe that the third AI reason is also true.  Given the way that large language models work, this is not a good assumption and could lead to users believing hallucinated information.

Another area for concern is that people may be more susceptible to rely on output from AI chat (without verifying it) in certain situations (e.g., time pressure, stress, low motivation). In our study, we observed one participant who clearly started relying more directly on the chat AI output (e.g. copying and pasting from the chat to their notes) when they were told they only had five minutes left to work on the task.

\textbf{Opportunities}
The first wave of generative AI + search systems essentially place a web search interface and an AI chat interface together (or near each other).  We see many opportunities for the interactive information retrieval community to explore novel ways to integrate these technologies and provide additional features to better meet users needs.  The first feature that our data suggests is that AI chat interfaces should provide links to sources that support the information they generate.  Some generative AI tools such as the new Bing provide citations and links.  However, the quality and accuracy of these source links has been questioned.  Given how LLMs generate text, providing definitive source links may be challenging.
The second opportunity we see is to use generative AI and LLMs as a \emph{component} of a conversational information interface, but not as a direct interface that users engage with.  Interacting directly with a LLM for information seeking is a tempting, but odd notion.  By design, LLMs generate text.  Additional layers and components are needed to filter, verify, attribute sources, etc.
Third, we see opportunities to provide users with interfaces backed by generative AI, but that are more tailored to specific information needs than just ``chat''.  For example, users may want to explore information in structured ways (e.g., through tables, charts, graphs).
Finally, we see opportunities to give users controls to customize the generative AI outputs to their needs (e.g., length, amount of factual vs creative information, amount of opinions).

\textbf{Educating users}
Our results also highlight several important points to help educate users about using chat AI systems. First, chat AI systems should be used as one of many sources and their output should be carefully verified against other authoritative sources.  Second, chat AI systems can be useful to provide overviews and perspectives, but users need to keep in mind that they may contain biases or incorrect information.  Third, users need to understand basic principles behind how generative AI systems create text. As part of this, users should be educated that just because some of the response is true does not mean all of it is. Finally, we think users should be educated to be cautious of relying on chat AI when under time pressure or other stressors.  These are situations where a person may be more likely to accept information that they have not had a chance to verify.

\section{Conclusion}
In this preprint paper, we report results from a small-scale user study to investigate how users incorporate AI chat into online search tasks.  We report on observed search behaviors and participants' likes/dislikes, trust, and mental models of the chat system. To the best of our knowledge, this is one of the first papers to report on the impacts of generative AI chat tools on search behaviors.

\section{Acknowledgements}
The lead author thanks the University of North Carolina School of Information and Library Science for a sabbatical during the Spring 2023 semester that made this research possible.